\pdfoutput=1
\documentclass[twocolumn,prb,superscriptaddress,floatfix]{revtex4-1}

\usepackage{graphicx}
\usepackage{amsmath,amssymb,bm}
\usepackage{dcolumn}
\usepackage{gensymb}
\usepackage{textcomp}
\usepackage{xcolor}
\usepackage{overpic}
\usepackage{hyperref}

\begin{document}

\title[Anisotropy Project]{Nearly Isotropic Magnon Transport in Epitaxial Lithium Aluminum Ferrite Thin Films}
\author{Yiming Li}
\email{yiming24@stanford.edu}
\affiliation{Department of Physics, Stanford University, Stanford, CA 94305, USA}
\affiliation{Geballe Laboratory for Advanced Materials, Stanford University, Stanford, CA 94305, USA}
\affiliation{Stanford Institute for Materials and Energy Sciences, SLAC National Accelerator Laboratory, Menlo Park, CA 94025, USA}

\author{Katya Mikhailova}
\affiliation{Department of Applied Physics, Stanford University, Stanford, CA 94305, USA}
\affiliation{Geballe Laboratory for Advanced Materials, Stanford University, Stanford, CA 94305, USA}
\affiliation{Stanford Institute for Materials and Energy Sciences, SLAC National Accelerator Laboratory, Menlo Park, CA 94025, USA}

\author{Lerato Takana}
\affiliation{Department of Applied Physics, Stanford University, Stanford, CA 94305, USA}
\affiliation{Geballe Laboratory for Advanced Materials, Stanford University, Stanford, CA 94305, USA}

\author{Daisy O'Mahoney}
\affiliation{Department of Materials Science and Engineering, Stanford University, Stanford, CA 94305, USA}
\affiliation{Geballe Laboratory for Advanced Materials, Stanford University, Stanford, CA 94305, USA}

\author{Sauviz P. Alaei}
\affiliation{Department of Physics, Stanford University, Stanford, CA 94305, USA}
\affiliation{Geballe Laboratory for Advanced Materials, Stanford University, Stanford, CA 94305, USA}

\author{Guanxiong Qu}
\affiliation{Department of Physics and Astronomy, University of California, Irvine, Irvine, CA 92697, USA}

\author{Dominic Petruzzi}
\affiliation{Department of Physics, Stanford University, Stanford, CA 94305, USA}
\affiliation{Geballe Laboratory for Advanced Materials, Stanford University, Stanford, CA 94305, USA}

\author{Samuel Crossley}
\affiliation{Geballe Laboratory for Advanced Materials, Stanford University, Stanford, CA 94305, USA}

\author{Harold Y. Hwang}
\affiliation{Department of Applied Physics, Stanford University, Stanford, CA 94305, USA}
\affiliation{Geballe Laboratory for Advanced Materials, Stanford University, Stanford, CA 94305, USA}
\affiliation{Stanford Institute for Materials and Energy Sciences, SLAC National Accelerator Laboratory, Menlo Park, CA 94025, USA}

\author{Ian R. Fisher}
\affiliation{Department of Applied Physics, Stanford University, Stanford, CA 94305, USA}
\affiliation{Geballe Laboratory for Advanced Materials, Stanford University, Stanford, CA 94305, USA}
\affiliation{Stanford Institute for Materials and Energy Sciences, SLAC National Accelerator Laboratory, Menlo Park, CA 94025, USA}

\author{Clare C. Yu}
\affiliation{Department of Physics and Astronomy, University of California, Irvine, Irvine, CA 92697, USA}

\author{Yuri Suzuki}
\affiliation{Department of Applied Physics, Stanford University, Stanford, CA 94305, USA}
\affiliation{Geballe Laboratory for Advanced Materials, Stanford University, Stanford, CA 94305, USA}
\affiliation{Stanford Institute for Materials and Energy Sciences, SLAC National Accelerator Laboratory, Menlo Park, CA 94025, USA}

\date{April 26, 2026}

\begin{abstract}
Low-loss magnetic insulating thin films are promising for information transport via magnons, where isotropic in-plane magnon propagation is desirable. We report nonlocal measurements of electrically and thermally generated magnons in epitaxial (001) lithium aluminum ferrite $\text{Li}_{0.5}\text{Al}_{0.7}\text{Fe}_{1.8}\text{O}_4$ thin films with pronounced fourfold in-plane magnetic anisotropy. By measuring the inverse spin Hall signal as a function of the magnon diffusion distance, we deduce magnon diffusion lengths that are nearly identical along the [100] and [110] directions at 250~K. This isotropy is consistent with a nearly isotropic exchange stiffness. These results highlight spinel ferrites as viable platforms for isotropic magnon transport.
\end{abstract}

\maketitle

Magnon transport in epitaxial thin films has been studied in a range of materials from conventional systems such as ferromagnets, antiferromagnets and ferrimagnets to more exotic materials such as non-collinear antiferromagnets, synthetic antiferromagnets and multiferroics.~\cite{sud2025overview} Among these, ferrimagnets are especially promising because of their fast spin dynamics similar to antiferromagnets and easy control in switching magnetization similar to ferromagnets.~\cite{kim2022ferrimagnetic} Moreover, ferrimagnetic insulators with low magnetic damping are excellent candidates for magnon propagation media as they give rise to minimal energy loss due to Joule heating compared to metals. The most well-studied ferrimagnetic insulators come from the family of garnet ferrites, including Y$_3$Fe$_5$O$_{12}$~\cite{arsad2023yig} and Tm$_3$Fe$_5$O$_{12}$.~\cite{avci2017tig} They exhibit Gilbert damping parameter as low as 2 x 10$^{-4}$ in thin film form~\cite{sun2012yig-damping, dallivy2013yig-damping, onbasli2014yig-damping, soumah2018yig-damping} and even lower with post-deposition annealing in oxygen.~\cite{hauser2016yig-annealing}

More recently, spinel ferrite thin films have been found to be a promising alternative to garnet ferrite films as they can be grown at much lower deposition temperatures, have lower external field requirements~\cite{emori2017nzafo,wisser2020mafo,zheng2020lfo} and have potential for integration with silicon.~\cite{ogale1998fe3o4,wang2019sto} Examples include $\text{Mg(Al,Fe)}_{2}\text{O}_4$ (MAFO) and $\text{Li}_{0.5}\text{(Al,Fe)}_{2.5}\text{O}_4$ (LAFO) thin films, \cite{emori2018mafo,zheng2023lafo,takana2025lafo} which have been made into heterostructure devices and demonstrated efficient spin transport.\cite{riddiford2019mafo,ren2023lafo,ren2025lafo,alaei2025lafo} In magnesium aluminum ferrite $\text{MgAl}_{0.5}\text{Fe}_{1.5}\text{O}_4$ thin films, previous work found a small fourfold in-plane magnetic anisotropy on the order of 10 mT according to static magnetization measurements.\cite{emori2018mafo} However, nonlocal spin transport measurements show a spin diffusion length 30\% greater for magnon propagation along the in-plane easier $\langle110\rangle$ axes compared to the harder $\langle100\rangle$ axes.~\cite{li2022mafo} It was proposed that the  anisotropic magnon transport was not a result of the magnetic anisotropy; instead, it could be explained by an anisotropy in the exchange stiffness.~\cite{li2022mafo}

To understand if the anisotropy in spin transport exists more broadly within spinel ferrite thin films, we studied epitaxial lithium aluminum ferrite $\text{Li}_{0.5}\text{Al}_{x}\text{Fe}_{2.5-x}\text{O}_4$ (LAFO) for $x=0.7$ thin films, which are spinel ferrites with larger in-plane magnetic anisotropy and even lower Gilbert damping than MAFO films. Epitaxial LAFO ($x=0.7$) films on MgAl$_2$O$_4$ substrates are under compressive epitaxial strain. They exhibit a magnetically hard axis out-of-plane and a fourfold in-plane magnetic anisotropy as do the epitaxial MAFO thin films, thus making them a suitable candidate to study spin transport for comparison with MAFO. Contrary to MAFO, we find a nearly isotropic spin diffusion length in the plane of the film. This phenomenon can be explained by nearly isotropic exchange energy in contrast to MAFO.~\cite{li2022mafo} This in-plane isotropic exchange is attributed to the tetragonally distorted crystal structure and weak spin-orbit coupling in the Fe$^{3+}$ cations in LAFO. In any case, the nearly isotropic spin diffusion lengths found in LAFO are promising for magnon waveguide or interconnect applications. 

Epitaxial LAFO ($x=0.7$) films were grown on $\text{MgAl}_2\text{O}_4$ (MAO) (001) substrates using pulsed laser deposition with a 248 nm KrF laser operating at a fluence of 2.8 J/cm$^2$. These epitaxial films were grown at substrate temperatures of 400--450 \degree C in $\text{O}_2$ atmospheres of 2.0 Pa, similar to previous studies on LAFO.~\cite{omahoney2023lafo, zheng2023lafo} X-ray diffraction measurements, performed on a Panalytical Empyrean X-ray diffractometer, indicate that the LAFO films grow epitaxially on MAO substrates under 2.2\% biaxial compressive epitaxial strain (See Supplement Section A). Films show excellent crystallinity with a typical rocking curve linewidth of $\Delta\omega = 0.0183\degree$ (See Supplementary Fig.~\ref{fig:struct-char} (a)--(b)). X-ray reflectivity (XRR) measurements show film thicknesses on the order of 10-15 nm for films grown for this study (Supplementary Fig.~\ref{fig:struct-char} (c)). Moreover, films are extremely smooth with a root-mean-square roughness of $\sim$0.2 nm as deduced from atomic force microscopy (Supplementary Fig.~\ref{fig:struct-char} (d)). 

\begin{figure}[ht]
    \centering
    \begin{overpic}[width=0.80\linewidth]{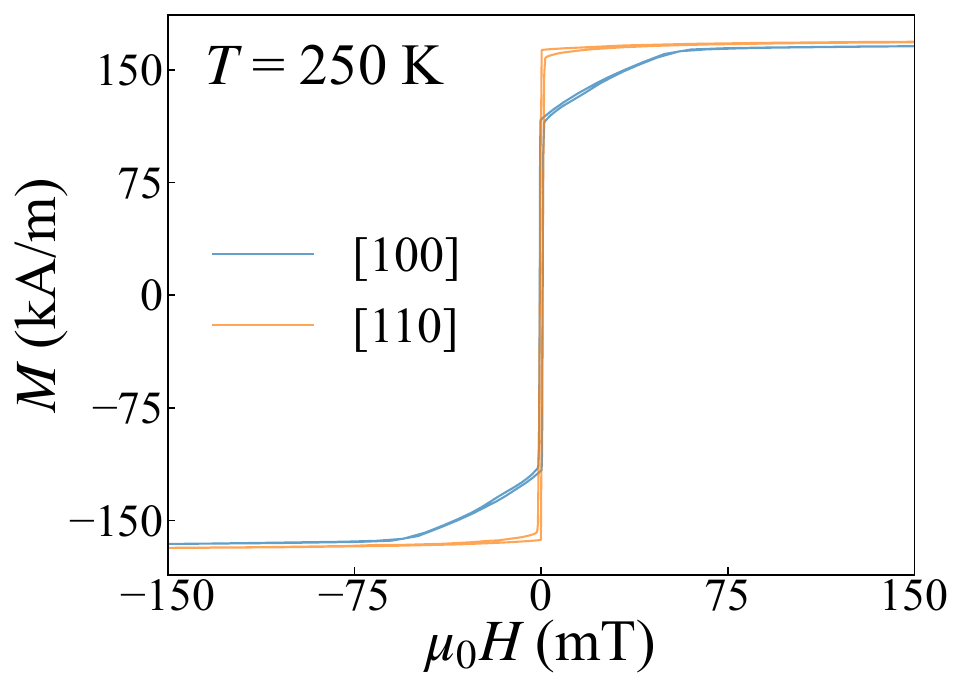}
        \put(58,26){\includegraphics[width=0.30\linewidth]{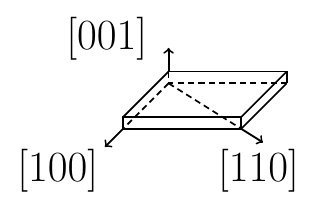}}
    \end{overpic}
    \caption{Field dependence of the magnetization for a (001) LAFO film on (001) MAO along the in-plane [100] and [110] directions at 250~K. (inset) Schematic diagram of the LAFO film showing the [001] out-of-plane axis and the [100] and [110] in-plane axes.} 
    \label{fig:MvH}
\end{figure}

We studied the magnetic anisotropy of our films via static magnetization measurements on a superconducting quantum interference device (SQUID) magnetometer. Epitaxial (001) LAFO films grown under compressive strain have a magnetically easier $\langle110\rangle$ family of axes and a magnetically harder $\langle100\rangle$ family of axes, with the out-of-plane [001] axis being the hardest.~\cite{omahoney2023lafo} Field dependence of the magnetization in our LAFO thin films shows square hysteresis loops with coercive fields on the order of 0.5--1.0 mT along the [110] in-plane axes at 250~K (Fig.~\ref{fig:MvH}). Along the magnetically harder [100] axis, the LAFO films exhibit fast domain switching followed by slow domain rotation until they saturate \cite{cullity2009magnetic}(Fig.~\ref{fig:MvH}). By fitting to the portion of the magnetization $M$ versus magnetic field $\mu_0H$ curve that corresponds to domain rotation (See Supplement Section B), we can extract the saturation magnetization $M_\text{s}=165.3\pm0.2$ kA/m and the saturation field $H_\text{s}=62\pm3$ mT. In the case of (001) oriented films of nominally cubic materials under biaxial epitaxial strain, the strain anisotropy does not affect the overall in-plane magnetic anisotropy. In addition, our sample is a 5 mm by 5 mm square, so shape anisotropy is negligible for in-plane measurements. What remains is the magnetocrystalline anisotropy, the energy of which can be computed as $K_1=\frac{-H_\text{s} M_\text{s}}{2}=(-5.1\pm0.3)\times10^3\text{ J}/\text{m}^3$.~\cite{cullity2009magnetic}

\begin{figure}[ht]
    \centering
    \begin{overpic}[width=0.48\linewidth]{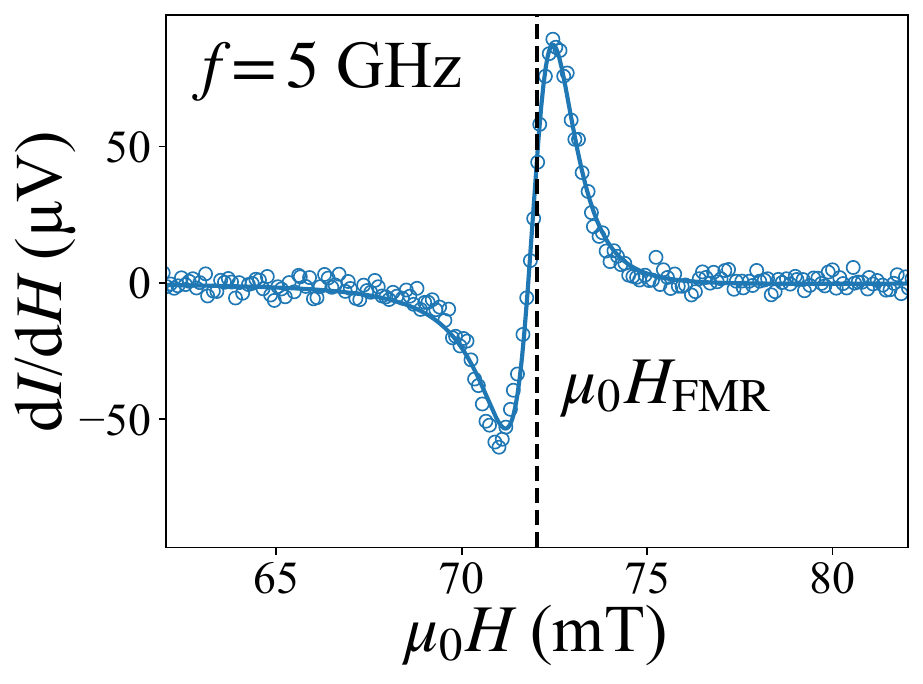}
      \put(0,70){(a)}
    \end{overpic}
    \hfill
    \begin{overpic}[width=0.48\linewidth]{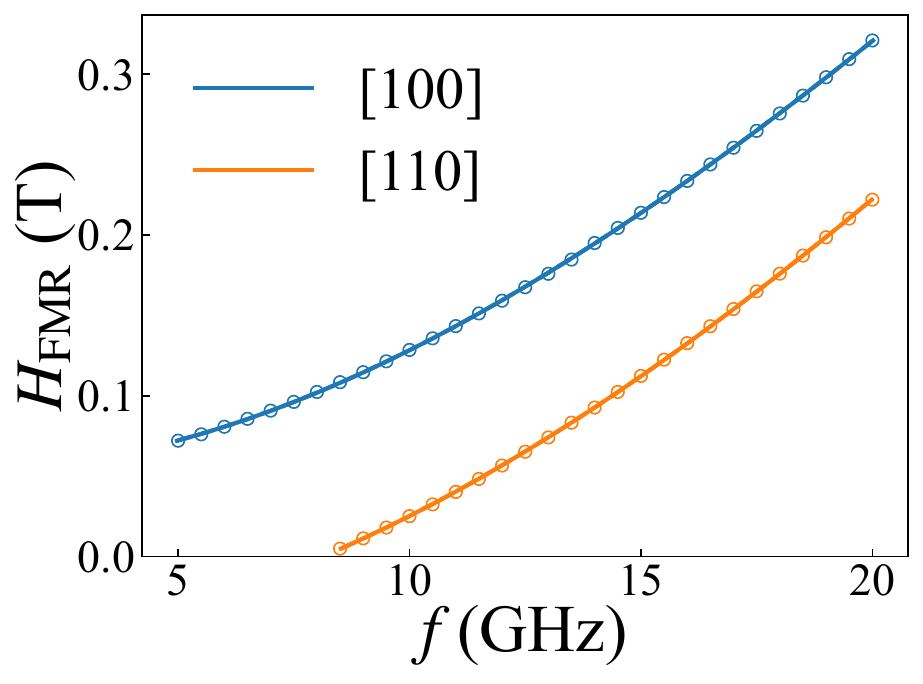}
      \put(0,70){(b)}
    \end{overpic}
    \caption{(a) Field dependence of the FMR signal at 5 GHz and 250~K. (b) Frequency dependence of $H_\text{FMR}$ at 250~K can be fit to the Kittel equation to deduce the Landé $g$-factor, effective magnetization $M_\text{eff}$ and in-plane anisotropy field $H_{4,\parallel}$ along the [100] and [110] axes.}
    \label{fig:FMR}
\end{figure}

Broadband ferromagnetic resonance (FMR) measurements demonstrate magnetic anisotropy values consistent with SQUID magnetometry measurements. FMR was measured in a coplanar waveguide geometry at frequencies between 5 GHz and 20 GHz at 250~K. The RF power transmitted through the sample was measured as a function of external magnetic field. As illustrated in Fig.~\ref{fig:FMR}(a), we extract the resonance field $H_\text{FMR}$ at 5 GHz by fitting the FMR spectrum to the generalized Lorentzian derivative.

By measuring the frequency dependence of the resonance field $H_\text{FMR}$, we can extract the Landé $g$-factor, effective magnetization $M_\text{eff}$ and fourfold in-plane anisotropy field $H_{4,\parallel}$ via the in-plane Kittel equation \cite{kittel1948fmr,farle1998fmr} (See Fig.~\ref{fig:FMR}(b)). Note that $M_\text{eff}=M_s-H_{2,\perp}$, where $M_s$ is the saturation magnetization, and $H_{2,\perp}$ is the uniaxial out-of-plane anisotropy field. Fitting along the [110] axis yields $g=1.98\pm0.02$, $\mu_0M_\text{eff}=1.70\pm0.03$ T and $\mu_0H_{4,\parallel}=-51.1\pm0.2$ mT. Since it is harder to fully saturate the sample along the [100] axis (See Supplement Section C), the best-fit parameters along the [100] axis have larger uncertainties. We do not expect $M_s$ or $H_{2,\perp}$ to vary in the (001) plane, so we can restrict $\mu_0M_\text{eff}$ to 1.65--1.75 T for fitting along the [110] axis. This yields $g=1.99\pm0.04$, $\mu_0M_\text{eff}=1.65\pm0.07$ T and $\mu_0H_{4,\parallel}=-53.2\pm0.2$ mT. The magnitude of fourfold in-plane anisotropy field $\lvert\mu_0H_{4,\parallel}\rvert$ of $\sim50$ mT is in quantitative agreement with the saturation field $\mu_0H_s=62\pm3$ mT from the static magnetometry measurements. 

Nonlocal magnon transport measurements, performed on LAFO films along the magnetically easier and harder in-plane directions, indicate nearly isotropic magnon transport. We used a nonlocal measurement geometry to probe magnon transport similar to previous studies on garnet ferrites.~\cite{cornelissen2015nonlocal,goennenwein2015nonlocal,shan2016nonlocal,shan2017nonlocal,oyanagi2020nonlocal} As shown in Fig.~\ref{fig:nonlocal-setup}, we deposited pairs of parallel Pt electrodes on the surface of LAFO films along the [100] and [110] directions and applied an AC current at $\omega/2\pi=9.8\text{ Hz}$ to one of the electrodes, creating a vertical spin current via the spin Hall effect (SHE). This leads to spin accumulation at the Pt/LAFO interface, thereby applying an anti-damping torque and exciting magnons in the LAFO film. These magnons propagate towards the second Pt electrode a distance $d$ away and are then detected electrically at this detector Pt electrode via the inverse spin Hall effect (ISHE). Additionally, the injection current generates a thermal gradient near the first electrode, producing magnons via the spin Seebeck effect (SSE). All magnons that arrive at the second electrode produce an electrical voltage via the ISHE. Since the SHE is proportional to the time-varying injection current $I(t)=I_0\cos(\omega t)$, we detected electrically generated magnons from the first harmonic voltage $V_{1\omega}$. The SSE is proportional to the square of the injection current $I^2(t)=I_0^2\cos^2(\omega t)$, so thermally injected magnons were detected from the second harmonic voltage $V_{2\omega}$. The nonlocal resistance is defined as $R_\text{SHE}=V_{1\omega}/I_0$ and $R_\text{SSE}=V_{2\omega}/I_0^2$, both of which are independent of the injection current $I_0$.

\begin{figure}[ht]
    \centering
    \begin{overpic}[width=0.48\linewidth]{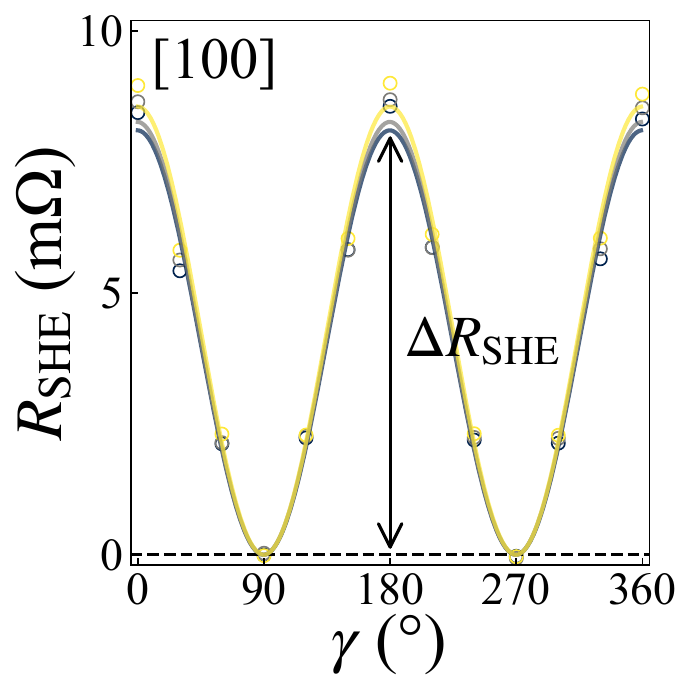}
      \put(0,90){(a)}
    \end{overpic}
    \hfill
    \begin{overpic}[width=0.48\linewidth]{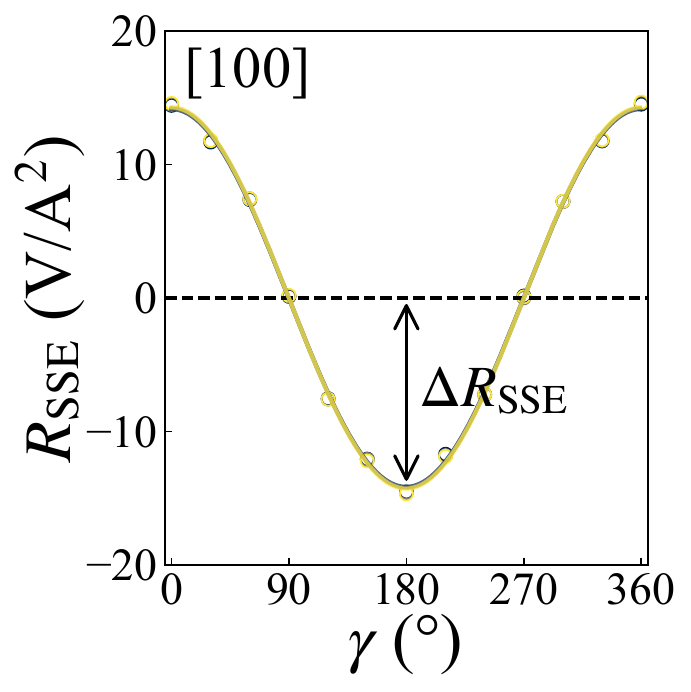}
        \put(0,90){(b)}
    \end{overpic}

    \begin{overpic}[width=0.48\linewidth]{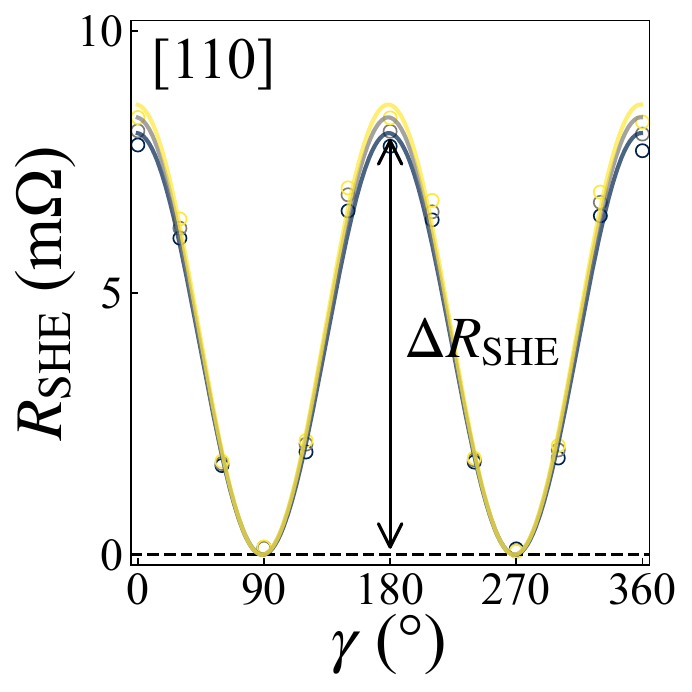}
      \put(0,90){(c)}
    \end{overpic}
    \hfill
    \begin{overpic}[width=0.48\linewidth]{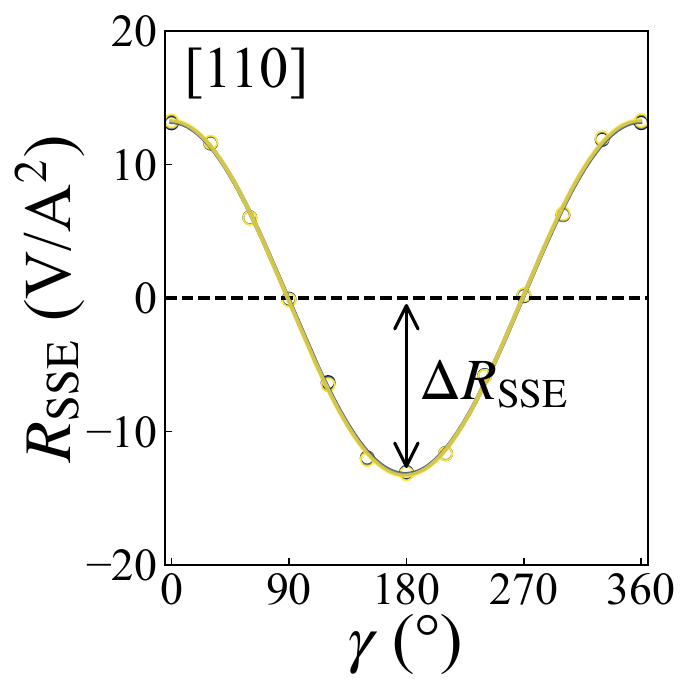}
        \put(0,90){(d)}
    \end{overpic}
    \caption{Nonlocal spin transport measurements along the [100] directions show (a) SHE nonlocal resistance $R_\text{SHE}$ and (b) SSE nonlocal resistance $R_\text{SSE}$ as a function of the angle $\gamma$ between the magnetization and the magnon propagation direction. Similar SHE and SSE measurements are shown for magnon propagation along the [110] direction in (c) and (d). These measurements were taken on a device with Pt electrodes a distance $d=0.5$ \textmu m apart at 250~K and 300 mT with 3 different injection currents $I_0$: 0.4 mA (yellow), 0.5 mA (grey) and 0.6 mA (blue).}
    \label{fig:R-vs-gamma} 
\end{figure}

We performed nonlocal measurements as a function of the in-plane orientation of the LAFO magnetization $\mathbf{M_s}$, which is saturated by an external magnetic field of 300 mT (See Fig.~\ref{fig:MvH}). The angle between magnetization $\mathbf{M_s}$ and the direction of magnon propagation is denoted by $\gamma$. Fig.~\ref{fig:R-vs-gamma} shows the $\gamma$ dependence of $R_\text{SHE}$ and $R_\text{SSE}$. $R_\text{SHE}$ has a $\cos^2{\gamma}$ dependence since the SHE and the ISHE each contribute a factor of $\cos{\gamma}$. $R_\text{SSE}$ only has a $\cos{\gamma}$ dependence from the ISHE because the SSE is independent of the direction of magnetization. By fitting to such $\gamma$ dependence, we extracted the amplitudes $\Delta R_\text{SHE}$ and $\Delta R_\text{SSE}$, respectively. We performed this measurement for magnons propagating along the [100] and [110] directions. For current values ranging from 0.4 to 0.6 mA, there is no significant difference in $\Delta R$. 

\begin{figure}[ht]
    \centering
    \includegraphics[width=0.80\linewidth]{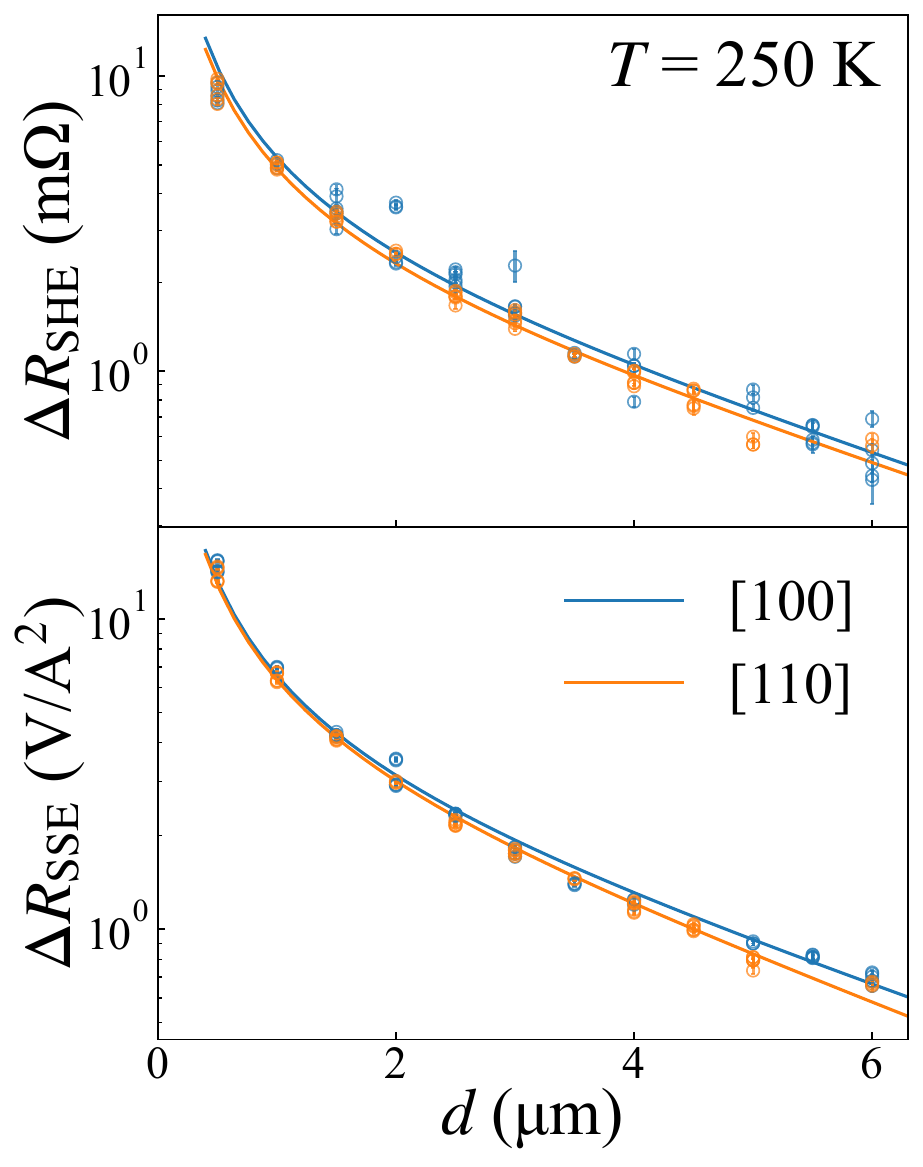}
    \caption{Amplitudes of the SHE nonlocal resistance $\Delta R_\text{SHE}$ (upper panel) and amplitudes of the SSE nonlocal resistance $\Delta R_\text{SSE}$ (lower panel) as a function of the magnon propagation distance $d$ along the [100] and [110] directions at 250~K. Measurements were taken with an injection charge current of $I_0=0.4$ mA, 0.5 mA and 0.6 mA on each device and one or two devices at the same distance $d$. A single best-fit curve is plotted for each orientation and for each type of nonlocal resistance.}
    \label{fig:A-vs-d}
\end{figure}

By measuring the nonlocal resistance on an array of parallel Pt electrodes with different separation distances $d$, we have deduced the spin diffusion length in LAFO films. Since the thickness of the LAFO film around 12 nm is much smaller than the distance $d$ between electrodes, and the distance $d$ is much smaller than the length of each electrode at 40 \textmu m, we can assume that magnons propagate in one dimension. The nonlocal resistance $\Delta R$ as a function of the separation distance $d$ is therefore described by the one-dimensional magnon diffusion equation:~\cite{cornelissen2015nonlocal}
\begin{equation}
    \Delta R \propto \frac{e^{-d/\lambda}}{1-e^{-2d/\lambda}},
    \label{eq:spin-diffusion}
\end{equation}
where $\lambda$ is known as the spin diffusion length. This equation encompasses both the diffusion-limited regime for small $d$ and the relaxation-limited regime for $d$ up to 3--5 times $\lambda$,~\cite{cornelissen2015nonlocal,shan2017nonlocal} consistent with the range of $d$ in our devices.

We measured the $d$ dependence of $\Delta R_\text{SHE}$ and $\Delta R_\text{SSE}$ as shown in Fig.~\ref{fig:A-vs-d}. By fitting the $d$ dependence of $\Delta R$ to Eq.~(\ref{eq:spin-diffusion}), we can estimate $\lambda$ along the [100] and [110] directions. Table~\ref{tab:spin-diffusion-length} lists the spin diffusion lengths $\lambda$ for SHE and SSE generated magnons along the [100] and [110] directions. At 250~K, $\lambda$ is independent of the magnon propagation direction ([100] vs [110]) and the generation mechanism (SHE vs SSE). Our results show that the in-plane spin diffusion length is similar along the [100] and [110] directions. The consistency between both types of magnons also indicates that secondary effects (such as the local generation of magnons due to the thermal gradient at the detector) does not strongly affect $\lambda_\text{SSE}$.~\cite{shan2017nonlocal,gao2022nonlocal}

\begin{table}[h]
    \centering
    \begin{tabular}{ccc}
    \hline
    Orientation & $\lambda_\text{SHE}$(\textmu m) & $\lambda_\text{SSE}$(\textmu m)\\ \hline
    [100] & 3.21\textpm0.21  & 3.25\textpm0.10 \\ \hline
    [110] & 3.24\textpm0.15  & 2.93\textpm0.07 \\ \hline
    \end{tabular}
    \caption{Spin-diffusion length $\lambda$ for SHE and SSE generated magnons propagating along the [100] and [110] axes at 250~K.}
    \label{tab:spin-diffusion-length}
\end{table}

Our nonlocal measurements provide compelling evidence that magnon transport in (001) LAFO films exhibit isotropic behavior with spin diffusion lengths on the order of 3 \textmu m. This result can be understood by considering the spin diffusion length $\lambda$ classically as:~\cite{zutic2004spintronics} $\lambda = \sqrt{D\tau_\text{nc}}$, where $D = \frac{1}{3}\langle v_g^2 \tau_\text{c}\rangle$ is the diffusion constant, $v_g$ is the magnon group velocity, and $\tau_\text{nc}$ and $\tau_\text{c}$ are the magnon non-conserving and magnon conserving relaxation times, respectively. If $\tau_\text{nc}$ and $\tau_\text{c}$ have any orientation dependence, then we would observe deviations from the expected behavior of $R_\text{SHE}\sim\cos^2(\gamma)$ and $R_\text{SSE}\sim\cos(\gamma)$. The lack of such orientation dependence means that both types of magnon relaxation time are isotropic, similar to MAFO.~\cite{li2022mafo}

Next, we consider the crystallographic orientation dependence of the magnon group velocity $v_g$ in LAFO. The crystallographic orientation dependence of $v_g$ of magnons is predominantly determined by the anisotropy of the exchange stiffness, which has a significant effect on magnons with larger wavenumber $k$.~\cite{li2022mafo} To estimate the anisotropy in the exchange stiffness, we consider the exchange interaction between neighboring Fe$^{3+}$ ions mediated by O$^{2-}$ anions. Since our LAFO films stabilize with a cubic crystal structure with slight tetragonal distortion, we expect little, if any, in-plane anisotropy in the exchange interaction, i.e. $J_{xx}=J_{yy}=J_{zz}$.~\cite{pryce1950perturbation, nagamiya1955anisotropy,belashchenko2004anisotropy,das2022anisotropy} The anisotropy of the exchange stiffness can therefore arise from either the spinel crystal structure or an anisotropic exchange interaction $J_{xy}$. However, the complex network of exchange paths in the LAFO spinel structure provides strong angular averaging of the exchange kernel, leading to nearly isotropic exchange stiffness. Similar angular averaging arising from complex exchange paths also underlies the nearly isotropic magnon dispersion observed in YIG.~\cite{CHEREPANOV199381,princep2017full} The anisotropy in exchange interaction can also arise when spin–orbit coupling (SOC) is incorporated into the virtual hopping processes underlying exchange. We consider the anisotropic exchange interaction that is symmetric in the spin indices, e.g., $J_{xy}\neq 0$. Its magnitude can be estimated as $\sim (\Lambda_\text{SOC}/\Delta)^2$, where $\Lambda_\text{SOC}$ is the strength of the spin-orbit coupling in Fe$^{3+}$ and $\Delta$ is the exchange splitting of the Fe$^{3+}$ ions.~\cite{van1951recent,nagamiya1955anisotropy} In our LAFO films, the SOC strength is on the order of 
$\Lambda_\text{SOC}\sim 70$ meV for an Fe$^{3+}$ ion,~\cite{Yeom_1996} and the energy scale of the exchange splitting $\Delta$ is on the order of $1$ eV for 3d transition metals.~\cite{Eastman_1980} Using these values, we estimate the symmetric exchange anisotropy to be less than 1\% of the isotropic exchange interaction. In addition, SOC causes the Landé $g$-factor to deviate from the free-electron value resulting in $(\Lambda_\text{SOC}/\Delta)\sim (g-2)$.~\cite{nagamiya1955anisotropy} Experimentally, we find $\lvert g-2\rvert\leq0.1$ (See Fig.~\ref{fig:FMR}(b) and Supplementary Fig.~\ref{fig:FMR-RT}(a)), which is also consistent with the above estimate.

Similar to LAFO, MAFO also has a cubic crystal structure and weak SOC from Fe$^{3+}$ ions, indicating a nearly isotropic exchange interaction. In contrast to the isotropic magnon transport in the plane in LAFO films, a 30\% anisotropy in the spin diffusion length has been reported in MAFO, which was attributed to an anisotropy in the exchange stiffness.~\cite{li2022mafo} Note that MAFO has a smaller in-plane magnetocrystalline anisotropy, with an in-plane anisotropy field of 10 mT at room temperature,~\cite{emori2018mafo} compared to LAFO of around 40 mT (See Supplementary Fig.~\ref{fig:FMR-RT}(a)). The SOC in MAFO is therefore on the same order as LAFO, if not smaller. Therefore, SOC cannot explain the anisotropic exchange stiffness in MAFO. 

Nevertheless, MAFO has a smaller inhomogeneous broadening than LAFO, so the averaging of the exchange kernel may be less effective. The inhomogeneous broadening is usually quantified by the zero-frequency FMR linewidth $\Delta H_0$, which is consistently less than 0.2 mT in MAFO,~\cite{emori2018mafo} but higher in LAFO about 0.4--1.2 mT (See Supplementary Fig.~\ref{fig:FMR-RT}(b)). The increased inhomogeneous broadening in LAFO can be attributed to the volatile nature of the Li$^+$ ions, which may have a non-uniform composition throughout the sample and lead to magnetic inhomogeneities. The presence of fewer inhomogeneities in MAFO might provide less averaging of the exchange kernel, leading to anisotropic magnon dispersion despite nearly isotropic exchange interaction $J$. Further magnon dispersion experiments, such as Brillouin light scattering, along different crystallographic directions in MAFO might offer clues of the anisotropic magnon dispersion,~\cite{sandweg2010bls,tong2026direct} although such experiments are usually restricted to measuring magnons near the Brillouin zone center. Ab initio density functional theory modeling of the magnon dispersion could also shed light on the anisotropic exchange stiffness in MAFO.~\cite{tong2026direct}

We further highlight another distinction in the crystallographic dependence of magnon transport between LAFO and MAFO. Apart from an anisotropic spin diffusion length, MAFO also has an anisotropy in the injected magnon densities, for both SHE and SSE generated magnons. Notably, the nonlocal signal at small spacings $d$ between the two electrodes is consistently larger in the harder [100] direction compared to the easier [1$\overline{1}$0] direction in MAFO, indicating a higher injected magnon density along the harder [100] direction.~\cite{li2022mafo} This is in contrast to LAFO, where we find no systematic difference in the nonlocal signal along the [100] and [110] directions at all spacings, as shown in Fig.~\ref{fig:A-vs-d}. Indeed, at small $d$, the magnon densities generated by the two injection methods tend to converge, indicating that magnons are equally injected for electrodes oriented along the [100] and [110] directions. This crucial difference between MAFO and LAFO leads us to suspect that the magnon injection mechanisms are dependent on the crystallographic direction in MAFO, which may underlie the divergent behavior at small $d$ and might also lead to an anisotropic spin diffusion length.

In summary, we have demonstrated that magnon transport in epitaxial LAFO thin films with low magnetic loss is nearly isotropic with respect to the film plane, contrary to epitaxial MAFO thin films. Spin diffusion lengths of both types of magnons measured along the in-plane [100] axis are similar to those measured along [110] axis by nonlocal magnon transport technique, despite the in-plane magnetocrystalline anisotropy measured by both static SQUID magnetometry and dynamic broadband FMR. The nearly isotropic magnon transport in low damping LAFO thin films, while expected from a nearly isotropic exchange energy, is promising for incorporation into magnon waveguides and interconnects. 

\begin{acknowledgments}
This work was supported as part of the Center for Energy Efficient Magnonics, an Energy Frontier Research Center (CEEMag) funded by the U.S. Department of Energy, Office of Science, Basic Energy Sciences at SLAC National Laboratory under contract \# DE-AC02-76SF00515. Part of this work was performed at nano@stanford RRID:SCR\_026695. S.P.A. was supported by the National Science Foundation (NSF) under the Graduate Research Fellowship Program and the Air Force Office of Scientific Research, Grant No.FA9550-23-1-034. D.P. was supported by the Air Force Office of Scientific Research Award FA9550-24-1-0357. S.C. and I.R.F. were supported by the Department of Energy, Office of Basic Energy Sciences, under Contract No. DE-AC02-76SF00515. 
\end{acknowledgments}

\section*{Data Availability Statement}
The data that support the findings of this study are openly available in the Stanford Digital Repository at https://doi.org/10.25740/dn330pg4198.

\clearpage
\section*{Supplementary Information}
\setcounter{figure}{0}
\setcounter{table}{0}
\renewcommand{\thefigure}{S\arabic{figure}}
\renewcommand{\thetable}{S\arabic{table}}
\subsection{Structural Characterization}
\begin{figure}[ht]
    \centering
    \begin{overpic}[width=0.48\linewidth]{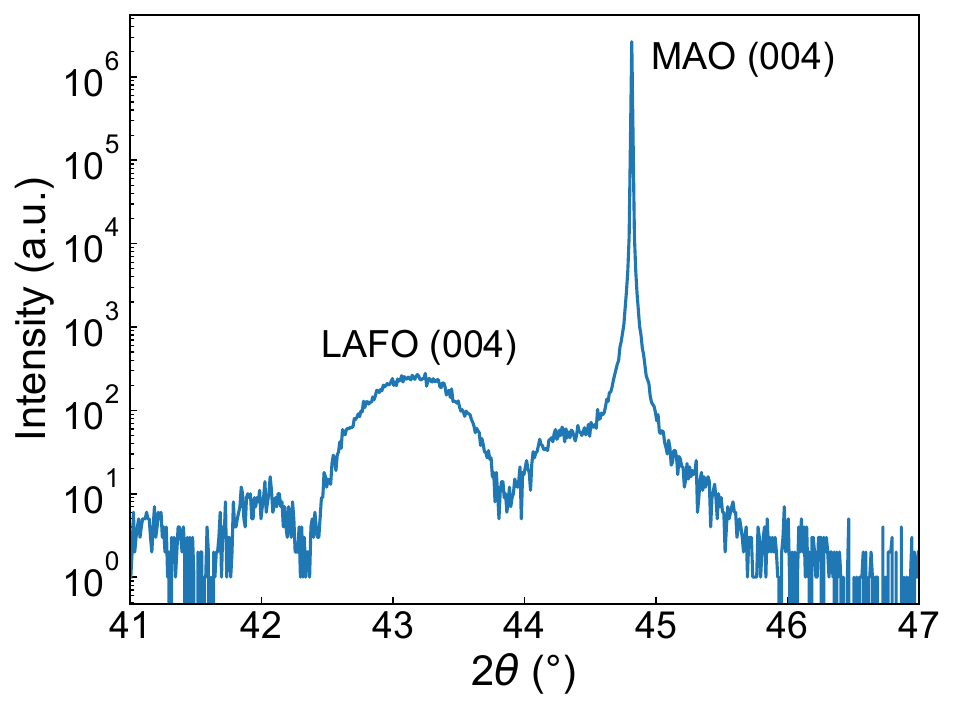}
        \put(0,68){(a)}
    \end{overpic}
    \hfill
    \begin{overpic}[width=0.48\linewidth]{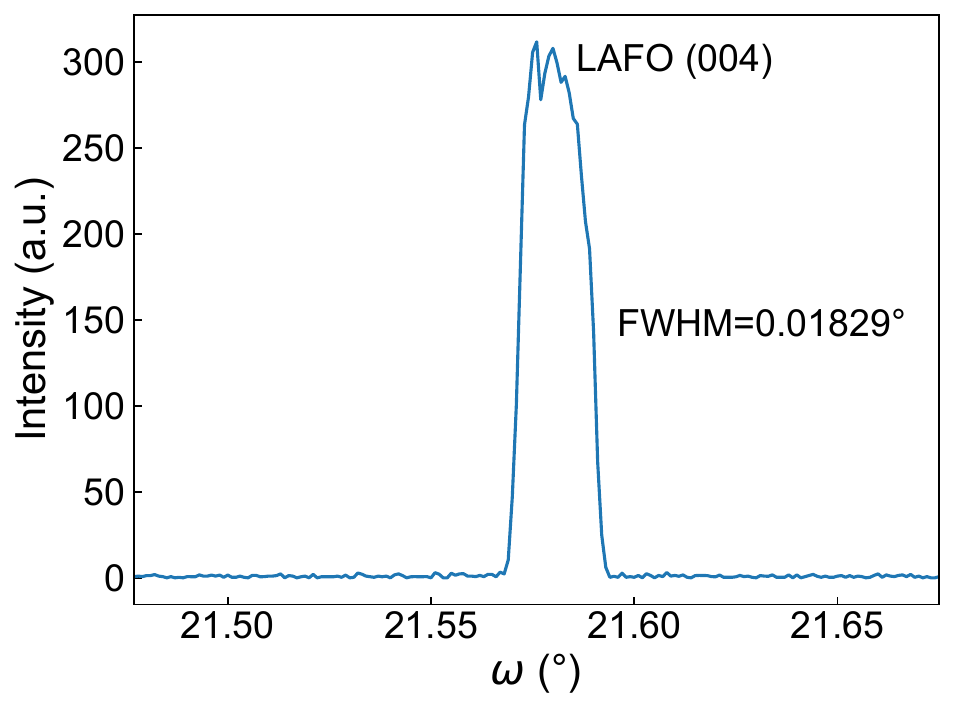}
        \put(0,68){(b)}
    \end{overpic}
    
    \begin{overpic}[width=0.48\linewidth]{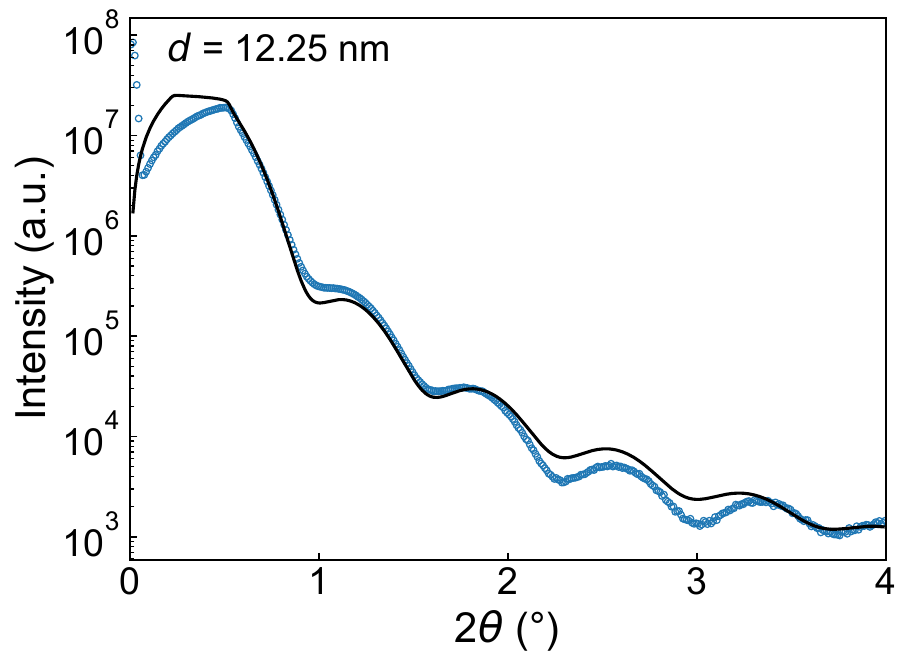}
        \put(0,68){(c)}
    \end{overpic}
    \hfill
    \begin{overpic}[width=0.48\linewidth]{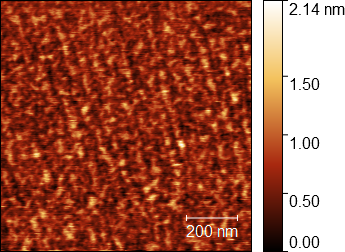}
        \put(0,68){\textcolor{white}{(d)}}
    \end{overpic}
    %captionsetup{justification=raggedright,singlelinecheck=false}
    \caption{(a) XRD data showing the MAO (004) peak and the LAFO (004) peak to its left. (b) Rocking curve at the LAFO (004) peak showing a FWHM of 0.01829\degree. (c) XRR result (blue) and the best-fit curve (black) of a 12.25 nm LAFO film. (d) AFM image of a 1 \textmu m by 1 \textmu m region on the LAFO surface showing a RMS roughness of 0.234 nm.}
    \label{fig:struct-char}
\end{figure}

X-ray diffraction (XRD) measurements indicate that our epitaxial $\text{Li}_{0.5}\text{Al}_{0.7}\text{Fe}_{1.8}\text{O}_4$ (LAFO) thin films grow with excellent crystallinity and minimal defects. $2\theta$-$\omega$ measurements were taken near the MAO (004) substrate peak. As shown in Fig. \ref{fig:struct-char}(a), XRD scans show the (004) family of peaks, and the LAFO (004) film peak appears to the left of the MAO (004) substrate peak. The secondary peaks near the highly crystalline LAFO film (004) peak are Laue oscillations, which represent interference from the smooth film/substrate interface and film surface and indicate high crystallinity at such interfaces. The $\omega$ rocking curve measurement in Fig. \ref{fig:struct-char}(b) of the LAFO (004) peak reveals a full-width at half maximum (FWHM) of  0.01829\degree, indicating extremely low mosaic spread. For comparison, the rocking curve of the (004) MAO single crystal substrate peak yields a FWHM of 0.01653\degree, which is very similar to that of the LAFO film. This indicates that the LAFO film has crystalline quality similar to that of the single crystal substrate.

We can estimate the epitaxial strain based on the XRD data. The location of the MAO (004) substrate peak in the XRD data is calibrated against the lattice parameter of 0.808 nm for bulk MAO in the literature \cite{yamanaka1983mao}. The location of the LAFO (004) peak indicates an out-of-plane lattice parameter $c_\perp$ of 0.836 nm for LAFO. Since LAFO ($x=0.7$) has a bulk lattice constant of $a_0=c_0=0.826$ nm \cite{schulkes1963lafo}, this indicates a 1.2\% out-of-plane tensile strain. Moreover, LAFO films between 10 nm and 15 nm are typically under coherent strain \cite{omahoney2023lafo}, so we can assume that the in-plane lattice parameter $a_\parallel$ of LAFO takes on the MAO value of 0.808 nm. This corresponds to an in-plane compressive strain of 2.2\%.

X-ray reflectivity (XRR) and atomic force microscopy (AFM) measurements help us probe film thickness and surface morphology. The XRR spectrum in Fig. \ref{fig:struct-char}(c) indicates a film thickness of 12.25 nm. Additionally, AFM measurement of a 1 \textmu m by 1 \textmu m region on a typical 10--15nm thick LAFO surface shows a root-mean-square (RMS) roughness on the order of $\sim$0.2 nm. The RMS roughness is around a quarter of the lattice parameter.

\subsection{Calculation of the Saturation Magnetization and Saturation Field}
For a magnetic field $H$ parallel to the magnetically harder [100] axis, the magnetic domains first preferentially align along the easy [110] and [1$\bar{1}$0] axes. As the magnetic field increases, the domains rotate towards the [100] axis until $H$ reaches a threshold $H_\text{s}$, at which the magnetization saturates. The magnetization of the material as the domains rotate is implicitly described by \cite{cullity2009magnetic}
\begin{equation}
    \frac{H}{H_\text{s}} = 2\frac{M}{M_\text{s}} \left[\left(\frac{M}{M_\text{s}}\right)^2 - \frac{1}{2} \right],
    \label{eq:Kittel}
\end{equation}
where the saturation field $H_\text{s}$ and the saturation magnetization $M_\text{s}$ can be found by fitting to the SQUID data.

\subsection{Room Temperature Broadband FMR Measurements}
\begin{figure}[ht]
    \centering
    \begin{overpic}[width=0.48\linewidth]{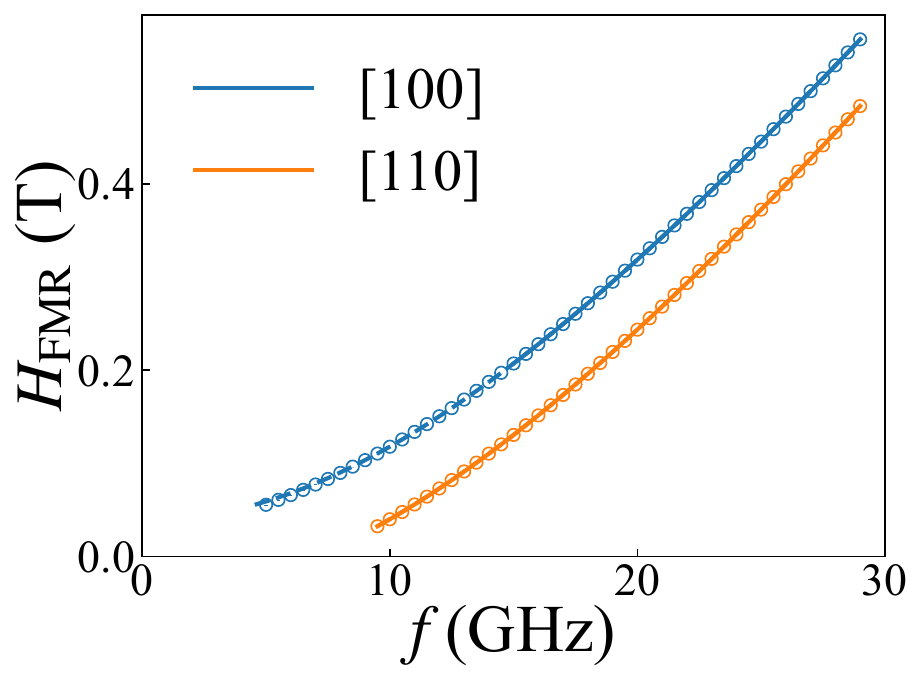}
        \put(0,72){(a)}
    \end{overpic}
    \hfill
    \begin{overpic}[width=0.48\linewidth]{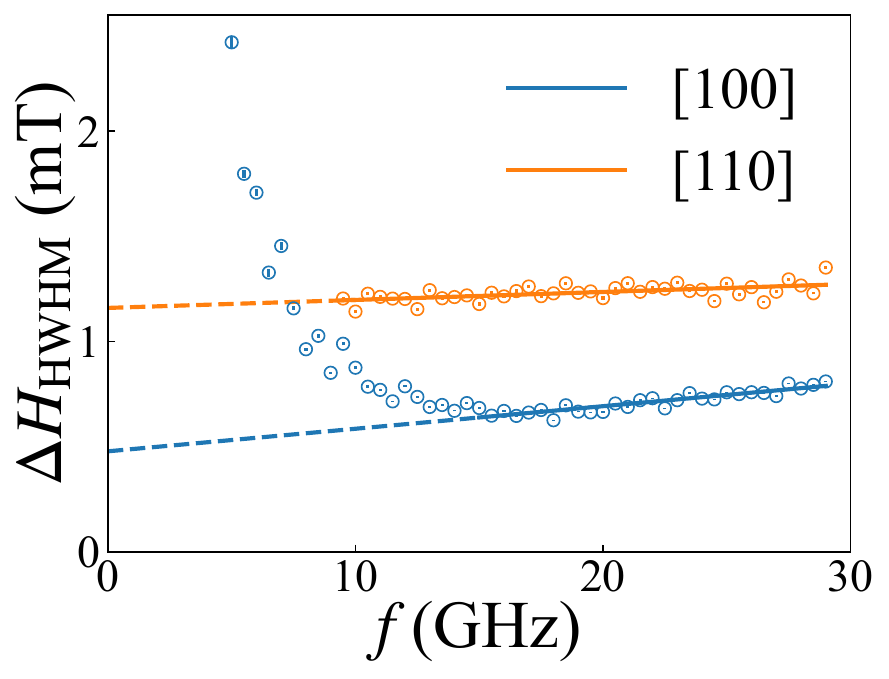}
        \put(0,72){(b)}
    \end{overpic}
    \caption{(a) Frequency dependence of the FMR resonance field $H_\text{FMR}$ along the [100] and [110] directions. (b) Frequency dependence of the FMR linewidth $\mu_0\Delta H_\text{HWHM}$ along the [100] and [110] directions. Fitting along the [100] direction starts at $f=15$ GHz. Solid curves show the fitting results, and dashed curves are interpolations of the best-fit curves to lower frequencies. Measurements are taken at room temperature.}
    \label{fig:FMR-RT}
\end{figure}

Using a coplanar waveguide setup, we measured the frequency dependence of the broadband FMR resonance field $H_\text{FMR}$ at room temperature, as shown in Fig.~\ref{fig:FMR-RT}(a). By fitting the resonance field along the [100] direction to the Kittel equation~\cite{kittel1948fmr,farle1998fmr}, we obtained a Landé factor of $g=2.081\pm0.004$, effective magnetization of $M_\text{eff}=1.402\pm0.008$ and fourfold in-plane anisotropy field of $H_{4,\parallel}=-38.40\pm0.27$ mT. Note that along the [100] direction, we only fit to data starting from 15 GHz because of the low-field loss in the linewidth (described below). Data along the [110] direction gives similar best-fit parameters, with $g=2.047\pm0.002$, $M_\text{eff}=1.491\pm0.003$ T and $H_{4,\parallel}=-40.64\pm0.06$ mT.

The frequency dependence of the FMR linewidth $\mu_0\Delta H_\text{HWHM}$ is a linear function described by \cite{farle1998fmr},
\begin{equation}
    \Delta H_\text{HWHM} = \Delta H_0 + \frac{h}{g\mu_0\mu_B}\alpha f,
    \label{eq:linewidth}
\end{equation}
where the Gilbert damping parameter $\alpha$ can be extracted from the slope and the inhomogeneous broadening $\Delta H_0$ from the zero-frequency linewidth. Data shown in Fig.~\ref{fig:FMR-RT}(b) yields damping of $\alpha=(3.1\pm0.3)\times10^{-4}$ and inhomogeneous broadening of $\mu_0\Delta H_0=0.48\pm0.02$ mT along the [100] direction. Note that the linewidth increases significantly for frequencies below 15 GHz. This so-called low-field loss has been attributed to incomplete saturation of the sample along the magnetically harder axis~\cite{nembach2011perpendicular} and has been similarly observed in Li$_{0.5}$Fe$_{2.5}$O$_4$ and LAFO~\cite{zheng2020lfo,omahoney2023lafo}. Although the resonance field $H_\text{FMR}$ at these frequencies is well above the average anisotropy field of $\sim40$ mT, small regions of our LAFO thin films likely have very different anisotropy field, which makes saturating the entire sample difficult along the [100] direction. We therefore used data starting from 15 GHz to fit to both $H_\text{FMR}$ and $\Delta H_\text{HWHM}$ along the [100] direction.

The linewidth data along the [110] direction does not show low-field loss. This is expected because the entire sample can be easily saturated along the easier direction, despite slight grain misalignment. By fitting to Eq.~\ref{eq:linewidth}, we extract a damping of $\alpha=(1.1\pm0.2)\times10^{-4}$ and inhomogeneous broadening of $\mu_0\Delta H_0=1.16\pm0.02$ mT. Despite a lower damping parameter, the inhomogeneous broadening is much larger in the [110] direction compared to the [100] direction, which has been similarly noted in LAFO with a different composition~\cite{pal2026lowTfmr}. This difference cannot be explained by the contribution of in-plane mosaic spread to the linewidth, which is quantified by $\Delta H\approx \lvert\frac{dH_\text{FMR}}{d\phi}\rvert\Delta\phi$ and should be zero along the easier [110] and harder [100] axes~\cite{zakeri2007linewidth}. Nor could it be attributed to anisotropic two-magnon scattering, which could not simultaneously account for the larger offset and smaller slope of the [110] linewidth~\cite{zakeri2007linewidth,emori2016lsmo}.

\subsection{Nonlocal Measurement Setup}
\begin{figure}[ht]
    \centering
    \includegraphics[width=0.70\linewidth]{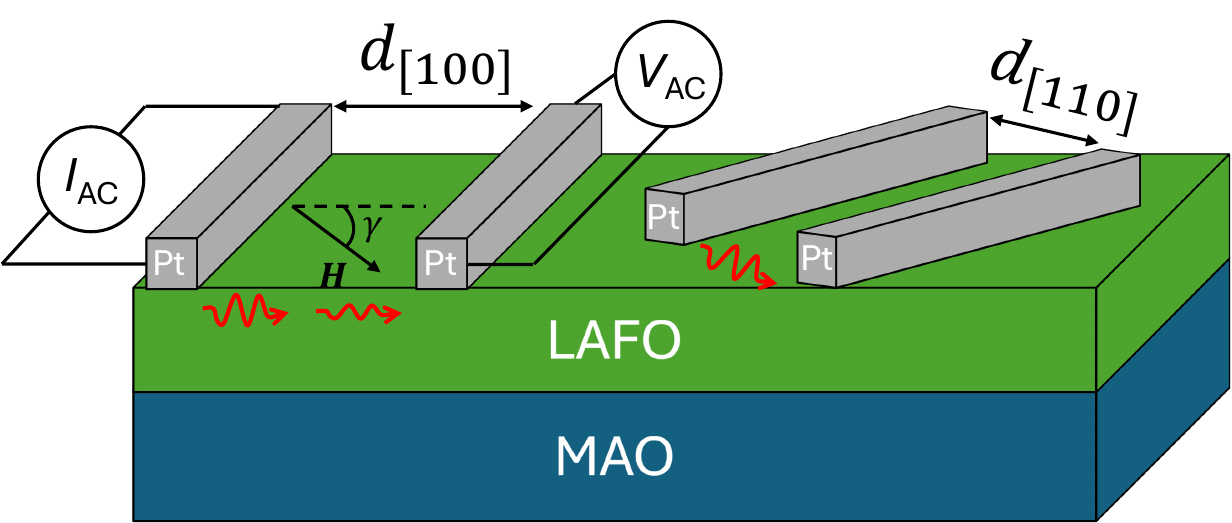}
    \caption{Nonlocal measurements were performed by injecting an AC current $I_\text{AC}$ into one Pt electrode and detecting the first and second harmonic voltage $V_\text{AC}$ at the opposite electrode. Each pair of electrodes can be oriented along the [100] direction or the [110] direction, and the variable distance between each pair is $d_\text{[100]}$ and $d_\text{[110]}$, respectively.}
    \label{fig:nonlocal-setup}
\end{figure}

As shown in Fig.~\ref{fig:nonlocal-setup}, we performed nonlocal measurements with pairs of parallel Pt electrodes deposited on the LAFO surface. The spin diffusion length can be extracted by measuring the nonlocal signal $V_\text{AC}$ on devices with different distances $d_\text{[100]}$ or $d_\text{[110]}$. By depositing devices with two different orientations, we can measure magnon diffusion along the [100] and [110] crystallographic directions.

\bibliography{references}% Produces the bibliography via BibTeX.

@article{omahoney2023lafo,
	author = {O'Mahoney, Daisy and Channa, Sanyum and Zheng, Xin Yu and Vailionis, Arturas and Shafer, Padraic and N'Diaye, Alpha T. and Klewe, Christoph and Suzuki, Yuri},	
	title = {Aluminum substitution in low damping epitaxial lithium ferrite films},
	journal = {Applied Physics Letters},
	year = {2023},
	volume = {123},
	number = {17},
	pages = {172405},
	doi = {10.1063/5.0163362}
}

@INBOOK{cullity2009magnetic,
   author       = "Cullity, B. D. and Graham, C. D.",
   title        = "Introduction to Magnetic Materials",
   chapter      = "7",
   pages        = "220",
   publisher    = "Wiley",
   year         = "2009",
   edition      = "2",
}

@article{li2022mafo,
    author = {Li, Ruofan and Li, Peng and Yi, Di and Riddiford, Lauren J. and Chai, Yahong and Suzuki, Yuri and Ralph, Daniel C. and Nan, Tianxiang},
    title = {Anisotropic Magnon Spin Transport in Ultrathin Spinel Ferrite Thin Films─Evidence for Anisotropy in Exchange Stiffness},
    journal = {Nano Letters},
    volume = {22},
    number = {3},
    pages = {1167-1173},
    year = {2022},
    doi = {10.1021/acs.nanolett.1c04332},
}

@article{belashchenko2004anisotropy,
    title = {Anisotropy of exchange stiffness and its effect on the properties of magnets},
    journal = {Journal of Magnetism and Magnetic Materials},
    volume = {270},
    number = {3},
    pages = {413-424},
    year = {2004},
    doi = {https://doi.org/10.1016/j.jmmm.2003.09.017},
    author = {K.D Belashchenko},
}

@article{das2022anisotropy,
	title = {Anisotropic long-range spin transport in canted antiferromagnetic orthoferrite {YFeO3}},
	volume = {13},
	doi = {10.1038/s41467-022-33520-5},
	number = {1},
	journal = {Nature Communications},
	publisher = {Nature Publishing Group},
	author = {Das, Shubhankar and Ross, A. and Ma, X. X. and Becker, S. and Schmitt, C. and van Duijn, F. and Galindez-Ruales, E. F. and Fuhrmann, F. and Syskaki, M.-A. and Ebels, U. and Baltz, V. and Barra, A.-L. and Chen, H. Y. and Jakob, G. and Cao, S. X. and Sinova, J. and Gomonay, O. and Lebrun, R. and Kläui, M.},
	month = oct,
	year = {2022},
	pages = {6140}
}

@article{pryce1950perturbation,
    doi = {10.1088/0370-1298/63/1/304},
    year = {1950},
    month = {jan},
    publisher = {},
    volume = {63},
    number = {1},
    pages = {25},
    author = {M H L Pryce},
    title = {A Modified Perturbation Procedure for a Problem in Paramagnetism},
    journal = {Proceedings of the Physical Society. Section A}
}

@article{nagamiya1955anisotropy,
    author = {T. Nagamiya and K. Yosida and R. Kubo},
    title = {Antiferromagnetism},
    journal = {Advances in Physics},
    volume = {4},
    number = {13},
    pages = {1--112},
    year = {1955},
    publisher = {Taylor \& Francis},
    doi = {10.1080/00018735500101154},
    }

@article{emori2018mafo,
    author = {Emori, Satoru and Yi, Di and Crossley, Sam and Wisser, Jacob J. and Balakrishnan, Purnima P. and Khodadadi, Behrouz and Shafer, Padraic and Klewe, Christoph and N’Diaye, Alpha T. and Urwin, Brittany T. and Mahalingam, Krishnamurthy and Howe, Brandon M. and Hwang, Harold Y. and Arenholz, Elke and Suzuki, Yuri},
    title = {Ultralow Damping in Nanometer-Thick Epitaxial Spinel Ferrite Thin Films},
    journal = {Nano Letters},
    volume = {18},
    number = {7},
    pages = {4273-4278},
    year = {2018},
    doi = {10.1021/acs.nanolett.8b01261}
}

@article{emori2017nzafo,
    author = {Emori, Satoru and Gray, Benjamin A. and Jeon, Hyung-Min and Peoples, Joseph and Schmitt, Maxwell and Mahalingam, Krishnamurthy and Hill, Madelyn and McConney, Michael E. and Gray, Matthew T. and Alaan, Urusa S. and Bornstein, Alexander C. and Shafer, Padraic and N'Diaye, Alpha T. and Arenholz, Elke and Haugstad, Greg and Meng, Keng-Yuan and Yang, Fengyuan and Li, Dongyao and Mahat, Sushant and Cahill, David G. and Dhagat, Pallavi and Jander, Albrecht and Sun, Nian X. and Suzuki, Yuri and Howe, Brandon M.},
    title = {Coexistence of Low Damping and Strong Magnetoelastic Coupling in Epitaxial Spinel Ferrite Thin Films},
    journal = {Advanced Materials},
    volume = {29},
    number = {34},
    pages = {1701130},
    doi = {https://doi.org/10.1002/adma.201701130},
    year = {2017}
}

@article{emori2016lsmo,
  title = {Spin transport and dynamics in all-oxide perovskite ${\mathrm{La}}_{2/3}{\mathrm{Sr}}_{1/3}{\mathrm{MnO}}_{3}/{\mathrm{SrRuO}}_{3}$ bilayers probed by ferromagnetic resonance},
  author = {Emori, Satoru and Alaan, Urusa S. and Gray, Matthew T. and Sluka, Volker and Chen, Yizhang and Kent, Andrew D. and Suzuki, Y.},
  journal = {Phys. Rev. B},
  volume = {94},
  issue = {22},
  pages = {224423},
  numpages = {9},
  year = {2016},
  month = {Dec},
  publisher = {American Physical Society},
  doi = {10.1103/PhysRevB.94.224423},
}

@article{riddiford2019mafo,
    author = {Riddiford, Lauren J. and Wisser, Jacob J. and Emori, Satoru and Li, Peng and Roy, Debangsu and Cogulu, Egecan and van 't Erve, Olaf and Deng, Yong and Wang, Shan X. and Jonker, Berend T. and Kent, Andrew D. and Suzuki, Yuri},
    title = {Efficient spin current generation in low-damping Mg(Al, Fe)2O4 thin films},
    journal = {Applied Physics Letters},
    volume = {115},
    number = {12},
    pages = {122401},
    year = {2019},
    month = {09},
    doi = {10.1063/1.5119726}
}

@article{wisser2020mafo,
    author = {Wisser, Jacob J. and Riddiford, Lauren J. and Altman, Aaron and Li, Peng and Emori, Satoru and Shafer, Padraic and Klewe, Christoph and N'Diaye, Alpha T. and Arenholz, Elke and Suzuki, Yuri},
    title = {The role of iron in magnetic damping of Mg(Al,Fe)2O4 spinel ferrite thin films},
    journal = {Applied Physics Letters},
    volume = {116},
    number = {14},
    pages = {142406},
    year = {2020},
    month = {04},
    doi = {10.1063/5.0003628}
}

@article{zheng2020lfo,
    author = {Zheng, Xin Yu and Riddiford, Lauren J. and Wisser, Jacob J. and Emori, Satoru and Suzuki, Yuri},
    title = {Ultra-low magnetic damping in epitaxial Li0.5Fe2.5O4 thin films},
    journal = {Applied Physics Letters},
    volume = {117},
    number = {9},
    pages = {092407},
    year = {2020},
    month = {09},
    doi = {10.1063/5.0023077}
}

@article{zheng2023lafo,
	title = {Ultra-thin lithium aluminate spinel ferrite films with perpendicular magnetic anisotropy and low damping},
	volume = {14},
	doi = {10.1038/s41467-023-40733-9},
	number = {1},
	journal = {Nature Communications},
	publisher = {Nature Publishing Group},
	author = {Zheng, Xin Yu and Channa, Sanyum and Riddiford, Lauren J. and Wisser, Jacob J. and Mahalingam, Krishnamurthy and Bowers, Cynthia T. and McConney, Michael E. and N’Diaye, Alpha T. and Vailionis, Arturas and Cogulu, Egecan and Ren, Haowen and Galazka, Zbigniew and Kent, Andrew D. and Suzuki, Yuri},
	month = aug,
	year = {2023},
	pages = {4918}
}

@article{takana2025lafo,
    author = {Takana, Lerato and Channa, Sanyum and Zheng, Xin Yu and O'Mahoney, Daisy and Alaei, Sauviz and Li, Yuntian and Vailionis, Arturas and Shafer, Padraic and N'Diaye, Alpha T. and Klewe, Christoph and Fisher, Ian and Suzuki, Yuri},
    title = {Low damping (111) oriented lithium aluminum ferrite thin films for spin wave applications},
    journal = {Applied Physics Letters},
    volume = {127},
    number = {3},
    pages = {032406},
    year = {2025},
    month = {07},
    doi = {10.1063/5.0278599}
}

@article{ren2023lafo,
	title = {Hybrid spin {Hall} nano-oscillators based on ferromagnetic metal/ferrimagnetic insulator heterostructures},
	volume = {14},
	doi = {10.1038/s41467-023-37028-4},
	number = {1},
	journal = {Nature Communications},
	author = {Ren, Haowen and Zheng, Xin Yu and Channa, Sanyum and Wu, Guanzhong and O’Mahoney, Daisy A. and Suzuki, Yuri and Kent, Andrew D.},
	month = mar,
	year = {2023},
	pages = {1406},
}

@article{ren2025lafo,
    author = {Ren, Haowen and Lai, Ya-An and Channa, Sanyum and O’Mahoney, Daisy A. and Zheng, Xin Yu and Suzuki, Yuri and Kent, Andrew D.},
    title = {Electrical Detection of Spin-Hall-Induced Auto-oscillations in Lithium Aluminate Ferrite Thin Films},
    journal = {Nano Letters},
    volume = {25},
    number = {16},
    pages = {6399-6404},
    year = {2025},
    doi = {10.1021/acs.nanolett.4c06305}
}

@article{alaei2025lafo,
    author = {Alaei, S. P. and Raj, R. and Channa, S. and Takana, L. and O'Mahoney, D. and Zheng, X. Y. and Fleck, E. E. and Chen, T.-Y. and Galazka, Z. and Kent, A. D. and Mkhoyan, K. A. and Suzuki, Y.},
    title = {Efficient spin transport across a disordered interface in a low damping magnetic insulator/heavy metal bilayer},
    journal = {Applied Physics Letters},
    volume = {127},
    number = {26},
    pages = {262403},
    year = {2025},
    month = {12},
    doi = {10.1063/5.0294600}
}

@article{kittel1948fmr,
  title = {On the Theory of Ferromagnetic Resonance Absorption},
  author = {Kittel, Charles},
  journal = {Phys. Rev.},
  volume = {73},
  issue = {2},
  pages = {155--161},
  numpages = {0},
  year = {1948},
  month = {Jan},
  publisher = {American Physical Society},
  doi = {10.1103/PhysRev.73.155}
}

@article{farle1998fmr,
    doi = {10.1088/0034-4885/61/7/001},
    year = {1998},
    month = {jul},
    publisher = {},
    volume = {61},
    number = {7},
    pages = {755},
    author = {Michael Farle},
    title = {Ferromagnetic resonance of ultrathin metallic layers},
    journal = {Reports on Progress in Physics}
    }

@article{cornelissen2015nonlocal,
	title = {Long-distance transport of magnon spin information in a magnetic insulator at room temperature},
	volume = {11},
	doi = {10.1038/nphys3465},
	number = {12},
	journal = {Nature Physics},
	publisher = {Nature Publishing Group},
	author = {Cornelissen, L. J. and Liu, J. and Duine, R. A. and Youssef, J. Ben and van Wees, B. J.},
	month = dec,
	year = {2015},
	pages = {1022--1026}
}

@article{goennenwein2015nonlocal,
	title = {Non-local magnetoresistance in {YIG}/{Pt} nanostructures},
	volume = {107},
	doi = {10.1063/1.4935074},
	number = {17},
	journal = {Applied Physics Letters},
	author = {Goennenwein, Sebastian T. B. and Schlitz, Richard and Pernpeintner, Matthias and Ganzhorn, Kathrin and Althammer, Matthias and Gross, Rudolf and Huebl, Hans},
	month = oct,
	year = {2015},
	pages = {172405},
}

@article{shan2016nonlocal,
  title = {Influence of yttrium iron garnet thickness and heater opacity on the nonlocal transport of electrically and thermally excited magnons},
  author = {Shan, Juan and Cornelissen, Ludo J. and Vlietstra, Nynke and Ben Youssef, Jamal and Kuschel, Timo and Duine, Rembert A. and van Wees, Bart J.},
  journal = {Phys. Rev. B},
  volume = {94},
  issue = {17},
  pages = {174437},
  numpages = {15},
  year = {2016},
  month = {Nov},
  publisher = {American Physical Society},
  doi = {10.1103/PhysRevB.94.174437},
}

@article{shan2017nonlocal,
  title = {Criteria for accurate determination of the magnon relaxation length from the nonlocal spin Seebeck effect},
  author = {Shan, J. and Cornelissen, L. J. and Liu, J. and Youssef, J. Ben and Liang, L. and van Wees, B. J.},
  journal = {Phys. Rev. B},
  volume = {96},
  issue = {18},
  pages = {184427},
  numpages = {8},
  year = {2017},
  month = {Nov},
  publisher = {American Physical Society},
  doi = {10.1103/PhysRevB.96.184427},
}

@article{oyanagi2020nonlocal,
	title = {Magnetic field dependence of the nonlocal spin {Seebeck} effect in {Pt}/{YIG}/{Pt} systems at low temperatures},
	volume = {10},
	issn = {2158-3226},
	doi = {10.1063/1.5135944},
	number = {1},
	journal = {AIP Advances},
	author = {Oyanagi, Koichi and Kikkawa, Takashi and Saitoh, Eiji},
	month = jan,
	year = {2020},
	pages = {015031},
}

@article{gao2022nonlocal,
	title = {Magnon transport and thermoelectric effects in ultrathin Tm$_3$Fe$_5$O$_{12}$/Pt nonlocal devices},
	volume = {4},
	doi = {10.1103/PhysRevResearch.4.043214},
	number = {4},
	journal = {Physical Review Research},
	publisher = {American Physical Society},
	author = {Gao, Jialiang and Lambert, Charles-Henri and Schlitz, Richard and Fiebig, Manfred and Gambardella, Pietro and Vélez, Saül},
	month = dec,
	year = {2022},
	pages = {043214},
}

@article{yamanaka1983mao,
	title = {Order-disorder transition in {MgAl2O4} spinel at high temperatures up to 1700 °{C}},
	volume = {165},
	doi = {10.1524/zkri.1983.165.14.65},
	number = {1-4},
	journal = {Zeitschrift für Kristallographie - Crystalline Materials},
	publisher = {De Gruyter},
	author = {Yamanaka, T. and Takéuchi, Y.},
	month = dec,
	year = {1983},
	pages = {65--78}
}

@article{schulkes1963lafo,
    title = {Crystallographic and Magnetic properties of the systems lithium ferrite-aluminate and lithium ferrite-gallate},
    journal = {Journal of Physics and Chemistry of Solids},
    volume = {24},
    number = {12},
    pages = {1651-1655},
    year = {1963},
    issn = {0022-3697},
    doi = {https://doi.org/10.1016/0022-3697(63)90110-0},
    author = {J.A. Schulkes and G. Blasse}
}

@misc{sud2025overview,
	title = {Novel {Magnetic} {Materials} for {Spintronic} {Device} {Technology}},
	doi = {10.48550/arXiv.2501.19087},
	publisher = {arXiv},
	author = {Sud, A. and Kumar, A. and Cubukcu, M.},
	month = jan,
	year = {2025}
}

@article{kim2022ferrimagnetic,
	title = {Ferrimagnetic spintronics},
	volume = {21},
	doi = {10.1038/s41563-021-01139-4},
	number = {1},
	journal = {Nature Materials},
	publisher = {Nature Publishing Group},
	author = {Kim, Se Kwon and Beach, Geoffrey S. D. and Lee, Kyung-Jin and Ono, Teruo and Rasing, Theo and Yang, Hyunsoo},
	month = jan,
	year = {2022},
	pages = {24--34}
}

@article{arsad2023yig,
	title = {Recent {Advances} in {Yttrium} {Iron} {Garnet} {Films}: {Methodologies}, {Characterization}, {Properties}, {Applications}, and {Bibliometric} {Analysis} for {Future} {Research} {Directions}},
	volume = {13},
	shorttitle = {Recent {Advances} in {Yttrium} {Iron} {Garnet} {Films}},
	doi = {10.3390/app13021218},
	number = {2},
	journal = {Applied Sciences},
	publisher = {Multidisciplinary Digital Publishing Institute},
	author = {Arsad, Akmal Z. and Zuhdi, Ahmad Wafi Mahmood and Ibrahim, Noor Baa’yah and Hannan, Mahammad A.},
	month = jan,
	year = {2023},
	pages = {1218}
}

@article{avci2017tig,
	title = {Spin transport in as-grown and annealed thulium iron garnet/platinum bilayers with perpendicular magnetic anisotropy},
	volume = {95},
	doi = {10.1103/PhysRevB.95.115428},
	number = {11},
	journal = {Physical Review B},
	publisher = {American Physical Society},
	author = {Avci, Can Onur and Quindeau, Andy and Mann, Maxwell and Pai, Chi-Feng and Ross, Caroline A. and Beach, Geoffrey S. D.},
	month = mar,
	year = {2017},
	pages = {115428}
}

@article{dallivy2013yig-damping,
	title = {Inverse spin {Hall} effect in nanometer-thick yttrium iron garnet/{Pt} system},
	volume = {103},
	doi = {10.1063/1.4819157},
	number = {8},
	journal = {Applied Physics Letters},
	author = {d'Allivy Kelly, O. and Anane, A. and Bernard, R. and Ben Youssef, J. and Hahn, C. and Molpeceres, A H. and Carrétéro, C. and Jacquet, E. and Deranlot, C. and Bortolotti, P. and Lebourgeois, R. and Mage, J.-C. and de Loubens, G. and Klein, O. and Cros, V. and Fert, A.},
	month = aug,
	year = {2013},
	pages = {082408}
}

@article{hauser2016yig-annealing,
	title = {Yttrium {Iron} {Garnet} {Thin} {Films} with {Very} {Low} {Damping} {Obtained} by {Recrystallization} of {Amorphous} {Material}},
	volume = {6},
	doi = {10.1038/srep20827},
	number = {1},
	journal = {Scientific Reports},
	publisher = {Nature Publishing Group},
	author = {Hauser, Christoph and Richter, Tim and Homonnay, Nico and Eisenschmidt, Christian and Qaid, Mohammad and Deniz, Hakan and Hesse, Dietrich and Sawicki, Maciej and Ebbinghaus, Stefan G. and Schmidt, Georg},
	month = feb,
	year = {2016},
	pages = {20827}
}

@article{sun2012yig-damping,
	title = {Growth and ferromagnetic resonance properties of nanometer-thick yttrium iron garnet films},
	volume = {101},
	doi = {10.1063/1.4759039},
	number = {15},
	journal = {Applied Physics Letters},
	publisher = {AIP Publishing},
	author = {Sun, Yiyan and Song, Young-Yeal and Chang, Houchen and Kabatek, Michael and Jantz, Michael and Schneider, William and Wu, Mingzhong and Schultheiss, Helmut and Hoffmann, Axel},
	month = oct,
	year = {2012},
}

@article{onbasli2014yig-damping,
	title = {Pulsed laser deposition of epitaxial yttrium iron garnet films with low {Gilbert} damping and bulk-like magnetization},
	volume = {2},
	doi = {10.1063/1.4896936},
	number = {10},
	journal = {APL Materials},
	author = {Onbasli, M. C. and Kehlberger, A. and Kim, D. H. and Jakob, G. and Kläui, M. and Chumak, A. V. and Hillebrands, B. and Ross, C. A.},
	month = oct,
	year = {2014},
	pages = {106102},
}

@article{soumah2018yig-damping,
	title = {Ultra-low damping insulating magnetic thin films get perpendicular},
	volume = {9},
	doi = {10.1038/s41467-018-05732-1},
	number = {1},
	journal = {Nature Communications},
	publisher = {Nature Publishing Group},
	author = {Soumah, Lucile and Beaulieu, Nathan and Qassym, Lilia and Carrétéro, Cécile and Jacquet, Eric and Lebourgeois, Richard and Ben Youssef, Jamal and Bortolotti, Paolo and Cros, Vincent and Anane, Abdelmadjid},
	month = aug,
	year = {2018},
	pages = {3355},
}

@article{zakeri2007linewidth,
  title = {Spin dynamics in ferromagnets: Gilbert damping and two-magnon scattering},
  author = {Zakeri, Kh. and Lindner, J. and Barsukov, I. and Meckenstock, R. and Farle, M. and von H\"orsten, U. and Wende, H. and Keune, W. and Rocker, J. and Kalarickal, S. S. and Lenz, K. and Kuch, W. and Baberschke, K. and Frait, Z.},
  journal = {Phys. Rev. B},
  volume = {76},
  issue = {10},
  pages = {104416},
  numpages = {8},
  year = {2007},
  month = {Sep},
  publisher = {American Physical Society},
  doi = {10.1103/PhysRevB.76.104416},
}

@article{van1951recent,
  title={Recent developments in the theory of antiferromagnetism},
  author={Van Vleck, JH},
  journal={Journal de Physique et le Radium},
  volume={12},
  number={3},
  pages={262--274},
  year={1951}
}

@article{Yeom_1996,
  title = {EPR study of ${\mathrm{Fe}}^{3+}$ impurities in crystalline ${\mathrm{BiVO}}_{4}$},
  author = {Yeom, Tae Ho and Choh, Sung Ho and Du, Mao Lu and Jang, Min Su},
  journal = {Phys. Rev. B},
  volume = {53},
  issue = {6},
  pages = {3415--3421},
  numpages = {0},
  year = {1996},
  month = {Feb},
  publisher = {American Physical Society},
  doi = {10.1103/PhysRevB.53.3415},
}

@article{Eastman_1980,
  title = {Experimental Exchange-Split Energy-Band Dispersions for Fe, Co, and Ni},
  author = {Eastman, D. E. and Himpsel, F. J. and Knapp, J. A.},
  journal = {Phys. Rev. Lett.},
  volume = {44},
  issue = {2},
  pages = {95--98},
  numpages = {0},
  year = {1980},
  month = {Jan},
  publisher = {American Physical Society},
  doi = {10.1103/PhysRevLett.44.95},
}

@article{princep2017full,
  title={The full magnon spectrum of yttrium iron garnet},
  author={Princep, Andrew J and Ewings, Russell A and Ward, Simon and T{\'o}th, Sandor and Dubs, Carsten and Prabhakaran, Dharmalingam and Boothroyd, Andrew T},
  journal={npj Quantum Materials},
  volume={2},
  number={1},
  pages={63},
  year={2017},
  publisher={Nature Publishing Group UK London}
}

@article{zutic2004spintronics,
  title = {Spintronics: Fundamentals and applications},
  author = {\ifmmode \check{Z}\else \v{Z}\fi{}uti\ifmmode \acute{c}\else \'{c}\fi{}, Igor and Fabian, Jaroslav and Das Sarma, S.},
  journal = {Rev. Mod. Phys.},
  volume = {76},
  issue = {2},
  pages = {323--410},
  numpages = {0},
  year = {2004},
  month = {Apr},
  publisher = {American Physical Society},
  doi = {10.1103/RevModPhys.76.323},
}

@article{sandweg2010bls,
    author = {Sandweg, C. W. and Jungfleisch, M. B. and Vasyuchka, V. I. and Serga, A. A. and Clausen, P. and Schultheiss, H. and Hillebrands, B. and Kreisel, A. and Kopietz, P.},
    title = {Wide-range wavevector selectivity of magnon gases in Brillouin light scattering spectroscopy},
    journal = {Review of Scientific Instruments},
    volume = {81},
    number = {7},
    pages = {073902},
    year = {2010},
    month = {07},
    issn = {0034-6748},
    doi = {10.1063/1.3454918},
}

@article{nembach2011perpendicular,
  title = {Perpendicular ferromagnetic resonance measurements of damping and Land$\stackrel{\ifmmode \acute{}\else \'{}\fi{}}{\mathrm{e}}\phantom{\rule{4pt}{0ex}}g\ensuremath{-}$ factor in sputtered (Co${}_{2}$Mn)${}_{1\ensuremath{-}x}$Ge${}_{x}$ thin films},
  author = {Nembach, H. T. and Silva, T. J. and Shaw, J. M. and Schneider, M. L. and Carey, M. J. and Maat, S. and Childress, J. R.},
  journal = {Phys. Rev. B},
  volume = {84},
  issue = {5},
  pages = {054424},
  numpages = {15},
  year = {2011},
  month = {Aug},
  publisher = {American Physical Society},
  doi = {10.1103/PhysRevB.84.054424},
}

@misc{pal2026lowTfmr,
      title={Low-temperature spin dynamics in LAFO thin films: from cubic anisotropy to TLS-limited coherence}, 
      author={Srishti Pal and Guanxiong Qu and Hervé M. Carruzzo and Katya Mikhailova and Lerato Takana and Qin Xu and Yuri Suzuki and Clare C. Yu and Gregory D. Fuchs},
      year={2026},
      eprint={2602.05889},
      archivePrefix={arXiv},
      primaryClass={cond-mat.mtrl-sci},
      url={https://arxiv.org/abs/2602.05889}, 
}

@article{ogale1998fe3o4,
    author = {S.B. Ogale and K. Ghosh and S.P. Pai and M. Robson and Eric Li and I. Jin and R.L. Greene and R. Ramesh and T. Venkatesan and M. Johnson},
    title = {Fe3O4/SrTiO3/La0.7Sr0.3MnO3 heterostructure: growth and properties.},
    journal = {Materials Science and Engineering: B},
    volume = {56},
    number = {2},
    pages = {134-139},
    year = {1998},
    issn = {0921-5107},
    doi = {https://doi.org/10.1016/S0921-5107(98)00216-5},
}

@article{wang2019sto,
  title = {Epitaxial $\mathrm{SrTi}{\mathrm{O}}_{3}$ film on silicon with narrow rocking curve despite huge defect density},
  author = {Wang, Z. and Goodge, B. H. and Baek, D. J. and Zachman, M. J. and Huang, X. and Bai, X. and Brooks, C. M. and Paik, H. and Mei, A. B. and Brock, J. D. and Maria, J-P. and Kourkoutis, L. F. and Schlom, D. G.},
  journal = {Phys. Rev. Mater.},
  volume = {3},
  issue = {7},
  pages = {073403},
  numpages = {10},
  year = {2019},
  month = {Jul},
  publisher = {American Physical Society},
  doi = {10.1103/PhysRevMaterials.3.073403},
}

@article{CHEREPANOV199381,
title = {The saga of YIG: Spectra, thermodynamics, interaction and relaxation of magnons in a complex magnet},
journal = {Physics Reports},
volume = {229},
number = {3},
pages = {81-144},
year = {1993},
issn = {0370-1573},
doi = {https://doi.org/10.1016/0370-1573(93)90107-O},
url = {https://www.sciencedirect.com/science/article/pii/037015739390107O},
author = {Vladimir Cherepanov and Igor Kolokolov and Victor L'vov}
}

@article{tong2026direct,
author = {Tong, Junwei and Paudyal, Hari and Liu, Xiangcheng and Takana, Lerato and Mikhailova, Katya and Suzuki, Yuri and Paudyal, Durga and Li, Xiaoqin},
title = {Direct Observation of Tunable Magnons in Epitaxial Lithium Aluminum Ferrite Thin Films},
journal = {Nano Letters},
volume = {26},
number = {9},
pages = {3073-3079},
year = {2026},
doi = {10.1021/acs.nanolett.5c05922},
}

\end{document}